\documentclass[doublespacing]{elsart}
\usepackage[dvips]{color}

\usepackage{epsfig}
\usepackage{graphicx}
\usepackage{amssymb}
\usepackage{epstopdf}
\usepackage{cite}

\begin{document}
\begin{frontmatter}

\title{Formation of isolated islands by size-selected 
copper nanocluster deposition}
\author{Shyamal Mondal},
\author{B. Satpati,}
\author{S.R. Bhattacharyya\corauthref{cor}}
\ead{satyar.bhattacharyya@saha.ac.in}
\corauth[cor]{Corresponding author}

\address{Surface Physics Division, Saha Institute of Nuclear Physics, 1/AF Bidhan Nagar, Kolkata 700064, India}
\date{\today}

\begin{abstract}
\raggedright
Deposition of size-selected metal nanoclusters on a substrate with very low kinetic energy helps to keep the clusters intact with respect to their shape and size as compared to clusters in flight condition. Here we report formation of isolated monodispersed islands of copper of desired size on carbon film by deposition of size selected copper clusters ($\sim 3$ nm) in soft-landing method. Copper clusters were produced by a magnetron based gas aggregation type source equipped with a quardrupole mass filter (QMF) to select size 
of clusters before landing. Transmission Electron Microscopy (TEM) study shows that diffusion of islands is very low.

\end{abstract}


\end{frontmatter}
\maketitle
\newpage
\raggedright
\section{Introduction}

Thin films of metal nanoclusters with a large fraction of atoms exposed to the surface are promising candidate from 
technological point of view. Their thermodynamic, optical, electronic, magnetic and mechanical properties are quite different from their bulk counterparts \cite{haber1, binns, srb, palmer03, kashtanov}. One of the most efficient ways to exhibit these properties \cite{harbich} is the preparation of 
monodispersed metal nanoclusters on a specific substrate. For this purpose size-selected nanoclusters with a low or zero deposition energy are ideal \cite{heiz} as was felt over the years. Isolated islands of nano dimension on surfaces is of immense importance in recent days \cite{jens}. Due to quantum confinement effect these dots exhibit many properties (mostly superior) which are far away from their bulk counterparts. These nanodimensional dots having sizes below their Bohr diameters are known as 'quantum dots' which are worth studying in every material from the fundamental point of view. Formation of isolated islands or dots of specific sizes is important not only from fundamental aspect  but also for many technological possibilities regarding memory, size and speed 
of future devices. Supported metal clusters of specific sizes are important to study for their catalytic properties \cite{tainoff}. Ionized cluster beam deposition (ICBD) is one of the methods to get nanosized dots on a substrate in a controlled way. To get continuous deposition of size selected clusters with controlled impact energy and with appreciable 
cluster flux magnetron based gas aggregation types nanocluster source introduced by Haberland et al.\cite{haber94, haber1}equipped with QMF is superior to other sources \cite{binns, majum}. Haberland et al. \cite{haber95} showed deposition of clusters in different energy regime yields different film morphologies. By molecular dynamic (MD) simulation they showed that 
low energy deposition keeps shapes and sizes of preformed 
clusters intact after deposition. Such low energy or near to zero energy deposition is known as soft-landing. In tune with MD simulations, we report here the formation of isolated nano-scale dots of the sizes comparable to the mass-selected clusters in flight by soft landing of ionized copper clusters on amorphous carbon films.

\section{Experimental}
A UHV (Ultra High Vacuum) deposition system (Oxford Applied Research, UK) with gas aggregation type magnetron based nanocluster source has been used for production of copper nanoclusters. After the clusters are formed in the aggregation region, they are passed through a quadrupole mass filter that is capable of analysing clusters to a desired size with a narrow window of mass spectrum.  For the present study, amorphous carbon films coated on a standard 300 mesh copper grid (Ni grid for energy dispersive x-ray spectroscopy study) were used 
as substrates for deposition. Argon and Helium gases were used for magnetron sputtering, aggregation and carrying the clusters to deposition chamber from the nanocluster source as no extraction arrangement is there in the system[described elsewhere]. Base pressure before deposition was $ <4\times10^{-9}$ mbar and during deposition pressure in deposition chamber and in magnetron chamber were$\sim 5.0\times10^{-4}$ mbar and  $\sim 2\times10^{-1}$ mbar respectively and cluster ion current was maintained at $\sim 2.8$ nA controlling different parameters which controls cluster production. Substrates were kept at zero Volt (grounded) to avoid any acceleration of the positively charged clusters before deposition. Substrate surface plane was normal to the incident cluster beam direction for all depositions. Clusters of diameter $\sim 3$ nm (mass $\sim 
77000$ amu i.e. $\sim 1200$ atoms, using Wigner-Seitz radius for Copper, $r_w$ = 0.147 nm, detailed calculations in \cite{srb}) were selected for 
deposition with QMF resolution below $35\%$. Care was taken to eliminate sources of kinetic energy of the clusters as much as possible. All the deposited samples on the same day of deposition were investigated using Transmission Electron Microscope (TEM) facility (FEI TECNAI G2 S-TWIN) operating at 200 kV. For the information about composition of deposited 
film, the Energy Dispersive X-ray analysis (EDX) of the films was performed in the same TEM using 200 kV electron beam.

\section{Results and Discussion}
Figure 1 shows the morphological properties of the deposited films. The TEM images presented in fig. 1(a-d) are taken for different time of deposition under same conditions (i.e, same ion current measured at QMF, magnetron power, parametersof mass filter etc.). The corresponding histograms of the TEM images are shown below them in fig 1 h(a-d). Each histogram shows a very narrow size distribution around 3 to 4 nm with an 
increase with time of deposition or with surface coverage. At lowest coverage or for 20 min of deposition, distribution is centred at $\sim 3$ nm of diameter which is the selected diameter of the clusters before deposition. 

\begin{figure}[h]
\centering
\includegraphics[width=15 cm]{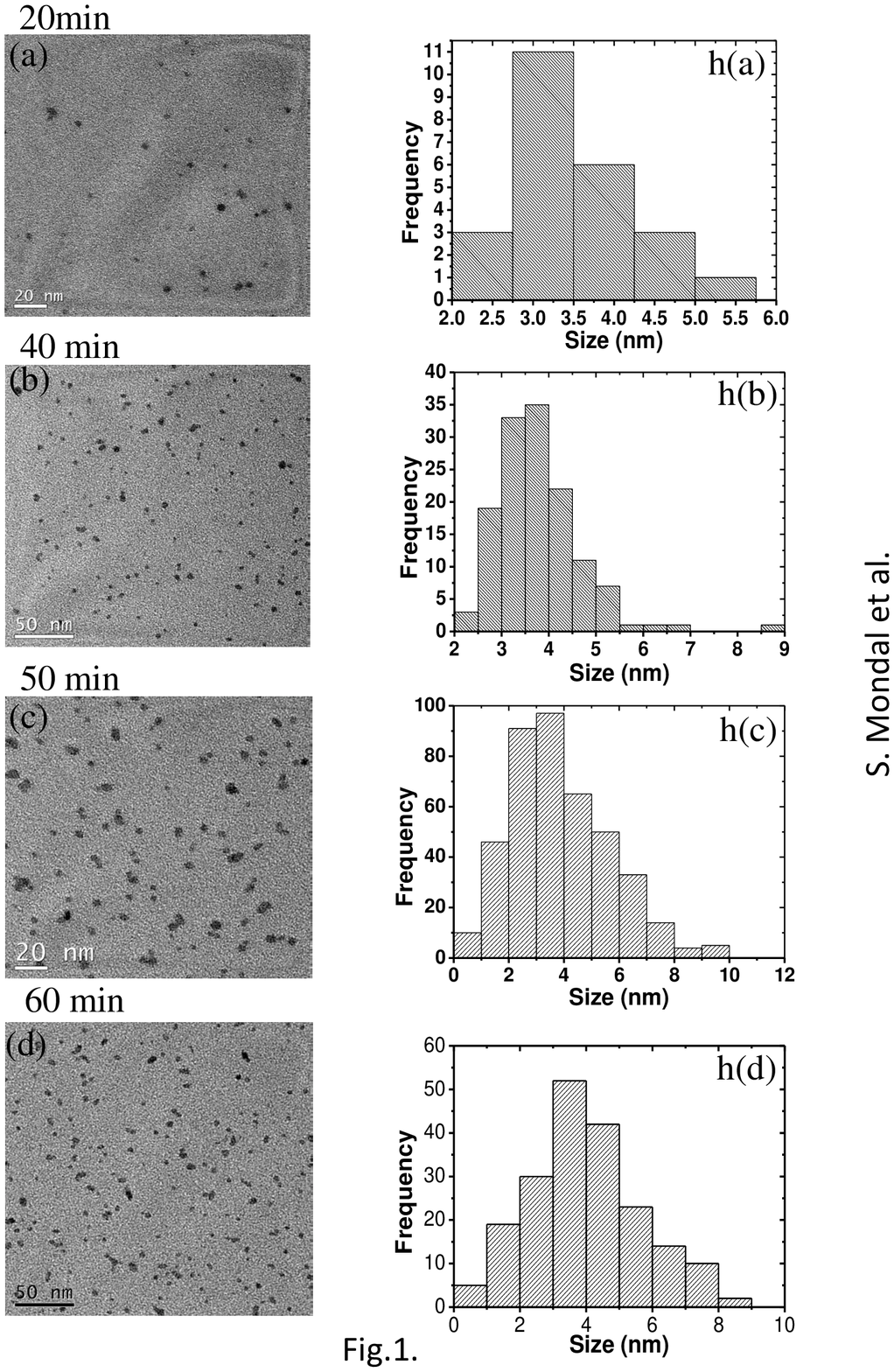}
\label{fig1}
\end{figure}

Figure 2(a) shows a TEM image from the 20 min deposited 
sample. A close observation shows that almost all the TEM
features are circular i.e, the clusters are spherical in 
shape. For longer time of deposition (e.g. about 1 hour), 
ramified islands are formed as shown in fig. 2(b). Some of the features are indicated here to show that  these features of TEM image of the 60 min deposited sample are not simple circular, instead they are ramified. For compositional information of films, the energy dispersive x-ray (EDX) analysis was performed using the attached EDX facility with TEM. Fig. 2(c) shows the EDX spectrum, depicting strong peaks of C, Ni and Cu. Island density vs deposition time plot in fig. 2(d) shows that initially island density increases linearly but it becomes saturated after certain value of coverage (here coverage is defined as percentage of surface area covered by clusters out of total surface area). This plot has been obtained calculating the island density for all the deposited sample in different time. Island density means number of island per unit area and when this obtained number is multiplied by the cluster projected area which supposed to be circular of diameter equal to the diameter the cluster then it becomes the number of island density per cluster site. This term certainly increases with time of deposition but if there is some saturation in the plot it gives the signature of surface diffusion and aggregation. The value of saturation island 
density measures the diffusion. If the value is very high then diffusion is not so high but there is a signature of diffusion as the island density is being saturated. In the present study island density was found to saturate at coverage below $5\%$ and it is very low\cite{jensen} than ideal substrate condition where the substrate has no defect sites and also the saturation island density value is quite high. Moreover the ramified shapes are not well branched or big in size. In the case of surface diffusion defect sites on the surface traps the clusters and their movement ceases \cite{francis,carroll} yielding to very low diffusion coefficient. Following Bardotti \textit{et al} \cite{bardotti} using the saturation island density in the equation$D=(0.41/N_{isl})^{1/\chi}F\pi d^4/16$, (where $D$ is surface diffusion coefficient, $N_{isl}$ is saturation island density per cluster site, $\chi$ = 0.336 \cite{bardotti}, $F$ is incident cluster flux and $d$ is cluster diameter), the surface diffusion coefficient in room temperature comes out to be 
equal to $\sim 20 ~nm^2/sec$, which is very low compared to other experiments where graphite substrate were used.

\begin{figure}[h]
\centering
\includegraphics[width=15 cm]{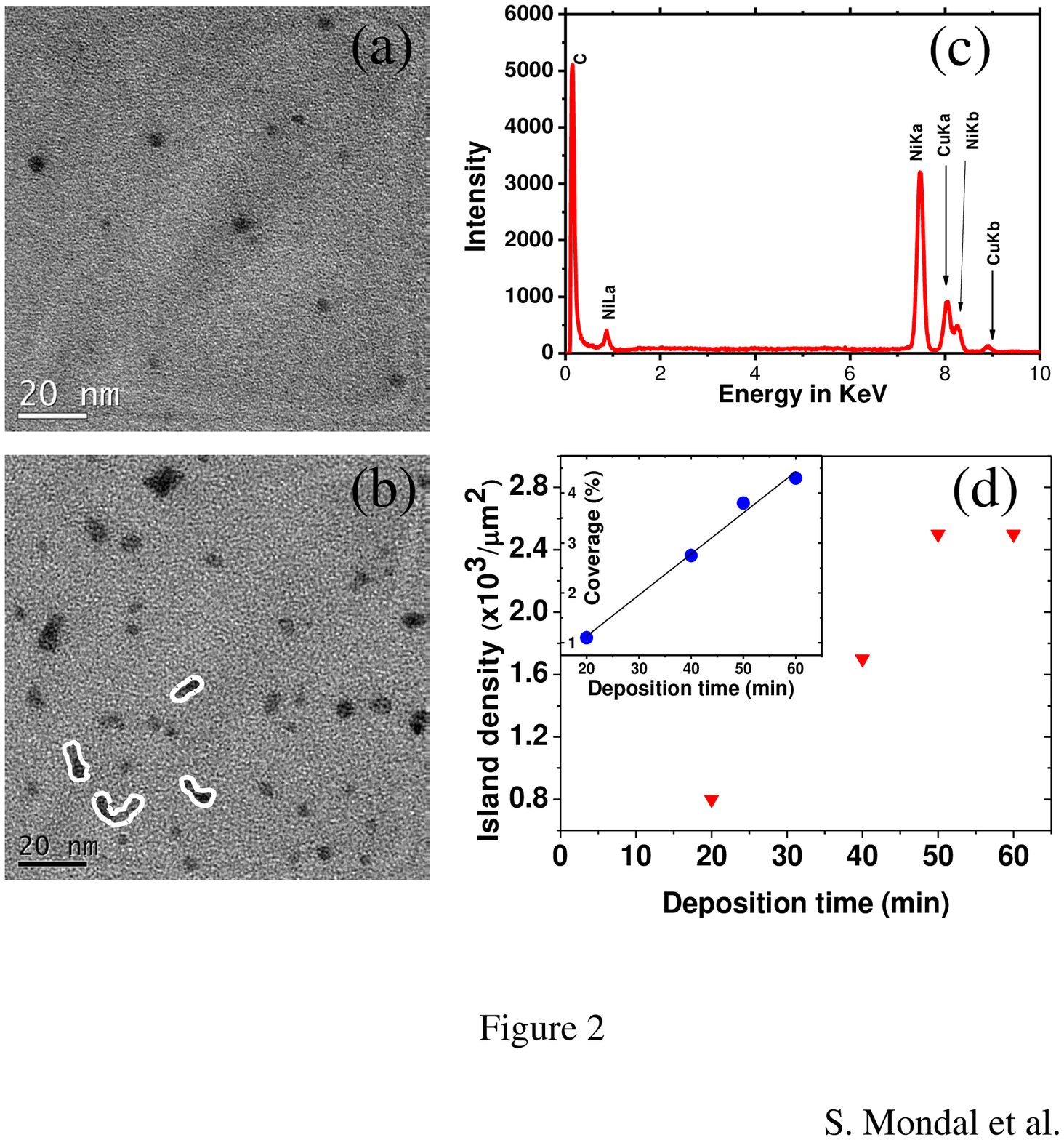}
\label{fig1}
\end{figure}

The in-set of 2(d) is derived from the average value of 
coverages at different time plotted against the time of 
deposition. This plot shows a linear monotonous increase in coverage with of time. This confirms the fact that 
deposited films are of submonolayer thickness and clusters are not falling on the previously deposited clusters or islands. The linear increase also gives the clear indication of the experimental fact that a constant flux has been maintained through all deposition experiments. Positively charged clusters which were deposited on the substrate $\sim 75$ cm away from the magnetron target were free from the negative bias applied to the target for sputtering and so as mentioned earlier that there is no other force to change the 
kinetic energy of the clusters except gas flow and so clusters attained very low kinetic energy near the soft landing.

Soft-landed clusters are not pinned to the substrate surface and so they diffuse on the substrate.  In deposition of cluster via soft-landing process the initial stage of growth of a film can be best described by the \textit{DDA model} \cite{jensen}, as deposition of incoming clusters takes place via diffusion and aggregation. Yoon  \textit{et al.} \cite{yoon} showed that a critical size of coalescence determines that the island will be formed by total coalescence or pure juxtaposition. Critical size of coalescence is that size of the clusters below which total coalescence occurs and clusters bigger than that do not coalesce but stick together to form ramified islands. 

In conclusion, isolated islands have been formed with 
controllable size by soft-landing Cu-nanoclusters on amorphous carbon film. The phenomenon is explained by different models of cluster deposition. In the present case for large copper clusters (1200 atoms/cluster), the diffusion is very slow as compared to other experiments \cite{francis,bardotti}. As a result monodispersed film formation has been possible with nearly soft landing deposition as exhibited in the present study. Detailed investigations with other cluster materials and  substrates in this direction are in progress.

 \textbf{Figure captions:} \vspace{.2in}\\
\textbf{Fig.1.:} (Upper panel) shows the Transmission Electron 
Microscopy (TEM) images of the deposited Cu-nanocluster films 
on carbon-coated Ni grids at different times of deposition. 
The deposited times are 20, 40, 50 and 60 min presented in 
(a), (b), (c) and (d) respectively. The histograms of the 
images are shown in the lower panel as h(a) to h(d) 
corresponding to the images (a) to (d) respectively. \\  

\textbf{Fig.2.:} (a) The typical TEM image for 20 min 
deposition showing circular spot from spherical clusters, 
while for longer time (60 min) deposition ramified structures 
along with elongated features are seen in (b). In (c) Energy 
Dispersive X-ray (EDX) spectrum shows the composition of the 
film on TEM grid for 60 min deposition sample. The island 
density of each deposition time is presented in fig. 2(d). 
The in-set in this figure shows the coverage (\%) of the 
substrates by Cu-clusters as a function of deposition time. 


\begin{thebibliography}{50}

\bibitem{haber1} Haberland H, in: Springer Series in Chemical Physics 52 (Clusters of atoms and molecules I, Theory, Experiment and clusters of atoms) (1994) Chapter 1 and 3 p. 1 and 207
\bibitem{binns} Binns C, Surf Sci Rep 2001; 44: 1-49.
\bibitem{srb} Bhattacharyya S R, Datta D, Shyjumon I,  Smirnov B M, Chini T K, Ghose D et al, J. Phys. D: Appl. Phys. 2009; 42: 035306(1-9).
\bibitem{palmer03} Palmer R E, Pratontep S, Boyen H -G, Nature Mater. 2003; 2: 443-8.
\bibitem{kashtanov} Kashtanov P V, Hippler R, Smirnov B M, Bhattacharyya S R, Europhys. Letts. 2010; 90:16001(p1-p4).
\bibitem{harbich} Harbich W, in Meiwes-Broer K H (Ed.), 
\textit{Metal Clusters at Surfaces}, Springer, Berlin, 2000.
\bibitem{heiz} Heiz U, Bullock E L, J Mater Chem 2004; 14: 564-77.
\bibitem{jens} Jensen P, Rev. Mod. Phys 1999; 71: 1695-735. 
\bibitem{tainoff} Tainoff D, Bardotti L, Tournus F, Guiraud G, Boisron O, and M\'{e}linon P, J. Phys. Chem. C 2008; 112: 6842-49.
\bibitem{haber94} H. Haberland H, Mall M, Mosseler M,  Qiang Y, Rainers T, Turner Y, J Vac Sci Technol A  1994; 12: 2925-31.
\bibitem{majum} Majumdar A, K\"{o}pp D, Ganeva M, Datta D, Bhattacharyya S R, Hippler R, Rev. Sci. Instrum. 2009; 80: 095103(1-6).
\bibitem{haber95} Haberland H, Insepov Z, Moseler M, Phys. Rev. B  1995; 51: 11061-7.
\bibitem{francis} Francis GM, Goldby IM, Kuipers L, Issendorff B von, Palmer R E, J. Chem. Soc 1996; Dalton Trans: 665-71.
\bibitem{carroll} Carroll S J, Seeger K, Palmer R E, Appl. 
Phys. Lett 1997; 72: 305-7.
\bibitem{bardotti} Bardotti L, Jensen P, Hoareau A, Treilleux M, Cabaud B, Phys. Rev. Lett 1995; 74: 4694-7.
\bibitem{jensen} Jensen P, Barab\'{a}si A -L, Larralde H, 
Havlin S, Stanley H E, Phys. Rev. B 1994; 50: 15316-29.
\bibitem{yoon} Yoon B, Akulin V M, Cahuzac Ph , Carlier F, de Frutos M, Masson A, Mory C, Colliex C,  Br\'{e}chignac C, Surf. Sci 1999; 443: 76-88.

\end{thebibliography}
\end{document}